\documentclass[  ,final    ,numberedheadings ]{aipproc}

\layoutstyle{8x11double}

\begin{document}

\title{Solar-Terrestrial Simulations of CMEs with a Realistic Initiation Mechanism: Case Study for Active Region 10069}


\author{N.~Lugaz}{
address={Institute for Astronomy, University of Hawaii-Manoa, 2680 Woodlawn Dr., Honolulu 96822, USA}
}

\author{I.~I.~Roussev}{
  address={Institute for Astronomy, University of Hawaii-Manoa, 2680 Woodlawn Dr., Honolulu 96822, USA}
}

\author{I.~V.~Sokolov}{
  address={Center for Space Environment Modeling, University of Michigan, Hayward St., Ann Arbor 48418, USA }
}

\author{C.~Jacobs}{
  address={Centrum voor Plasma Astrofysica, KU Leuven, Belgium}
}

\begin{abstract}

Most simulations of coronal mass ejections (CMEs) to date either focus on the interplanetary propagation of a giant plasma ``blob'' without paying too much attention to its origin and to the formation process or they focus on the complex evolution of the coronal magnetic field due to (sub-)photospheric motions which result in an eruption. Here, we present global simulations of CMEs where coronal motions are used to produce a realistic evolution of the coronal magnetic field and cause an eruption. We focus on active region 10069, which produced a number of eruptions in late August 2002, including the August 24, 2002 CME -- a fast ($\sim$ 2000 km~s$^{-1}$) eruption originating from W81--, as well as a slower eruption on August 22, 2002 (originating from W62).
Using a three-dimensional magneto-hydrodynamic (MHD) simulation of these ejections with the Space Weather Modeling Framework (SWMF), we show how a realistic initiation mechanism enables us to study the deflection of the CME in the corona and in the heliosphere. Reconnection of the erupting magnetic field with that of neighboring streamers and active regions modify the solar connectivity of the field lines connecting to Earth and change the expected solar energetic particle fluxes. Comparing the results at 1 AU of our simulations with {\it in situ} observations by the {\it ACE} spacecraft, we propose an alternate solar origin for the shock wave observed at L1 on August 26.

\end{abstract}

\maketitle

\section{Introduction}
\subsection{Active Region 10069 and August Eruptions}
On August 24, 2002, active region (AR) 10069 was near the western limb of the Sun (W81) when it was the source of a fast ($\sim$ 1,900~km~s$^{-1}$) and wide coronal mass ejection (CME). This event has been well studied due to extensive observations remotely by LASCO and UVCS \citep[]{Raymond:2003} as well in-situ by the {\it Wind} and {\it ACE} spacecraft. Most importantly, it was associated with an intense Solar Energetic Particle (SEP) event \citep[e.g., see][]{Tylka:2006, Tan:2009}. The eruption was preceded 2 days earlier by a similar, but twice slower CME, which was also associated with a large SEP event but is not thought to have been associated with a shock wave at 1 AU. Particularly noteworthy is the fact that, at the start of the August 24 event, the proton flux at energies above 10 MeV was still ten times above the intensity before the August 22 CME. 

Western limb events such as the August 24, 2002 CME present a number of challenges for space weather prediction. Due to the Parker spiral, the Earth is magnetically connected with regions on the solar surface at around W55 (for a 400~km~s$^{-1}$ wind). Therefore, SEP events are preferentially associated with western-limb events. Events such as the August 24, 2002 CME present additional challenges since the SEP arrival time at Earth corresponds to a particle release height of less than 5~$R_\odot$ \citep[]{Tylka:2006}. To explain the observations, a shock wave must be formed low in the corona at the flanks of the CME or the magnetic field line connecting the Earth to the solar surface must significantly diverge from the nominal Parker spiral. Such large differences of up to 30$^\circ$ between flare sites and the footprint tracing of the magnetic field lines connecting the Earth to the Sun have been reported before \citep[]{Ippolito:2005}. 

Western limb events can be associated with ejecta and more often with shock waves at 1~AU \citep[]{Cane:1988}. It is believed that the August 24, 2002 CME was associated with a shock wave at 1~AU on August 26. This fact, again, seems to imply either a very large span of the shock wave, a large deflection of the CME, or a combination of both. However, in \citet{Lugaz:2009a}, we studied numerically the August 24, 2002 eruption, and our model could not explain the expansion and/or deflection required to associate the shock at 1~AU to the eruption on August 24. In this article, we investigate i) why the August 22 CME, which originated from the same region when it was closer to the disk center, was not associated with a shock at 1 AU, and ii) the importance of this preceding ejection on the transport of SEPs accelerated by the August 24 CME.

\subsection{Solar Wind and Solar Eruption Models}

\begin{figure}[t]
\begin{minipage}[]{1.0\linewidth}
\begin{center}
{\includegraphics*[width=7.1cm]{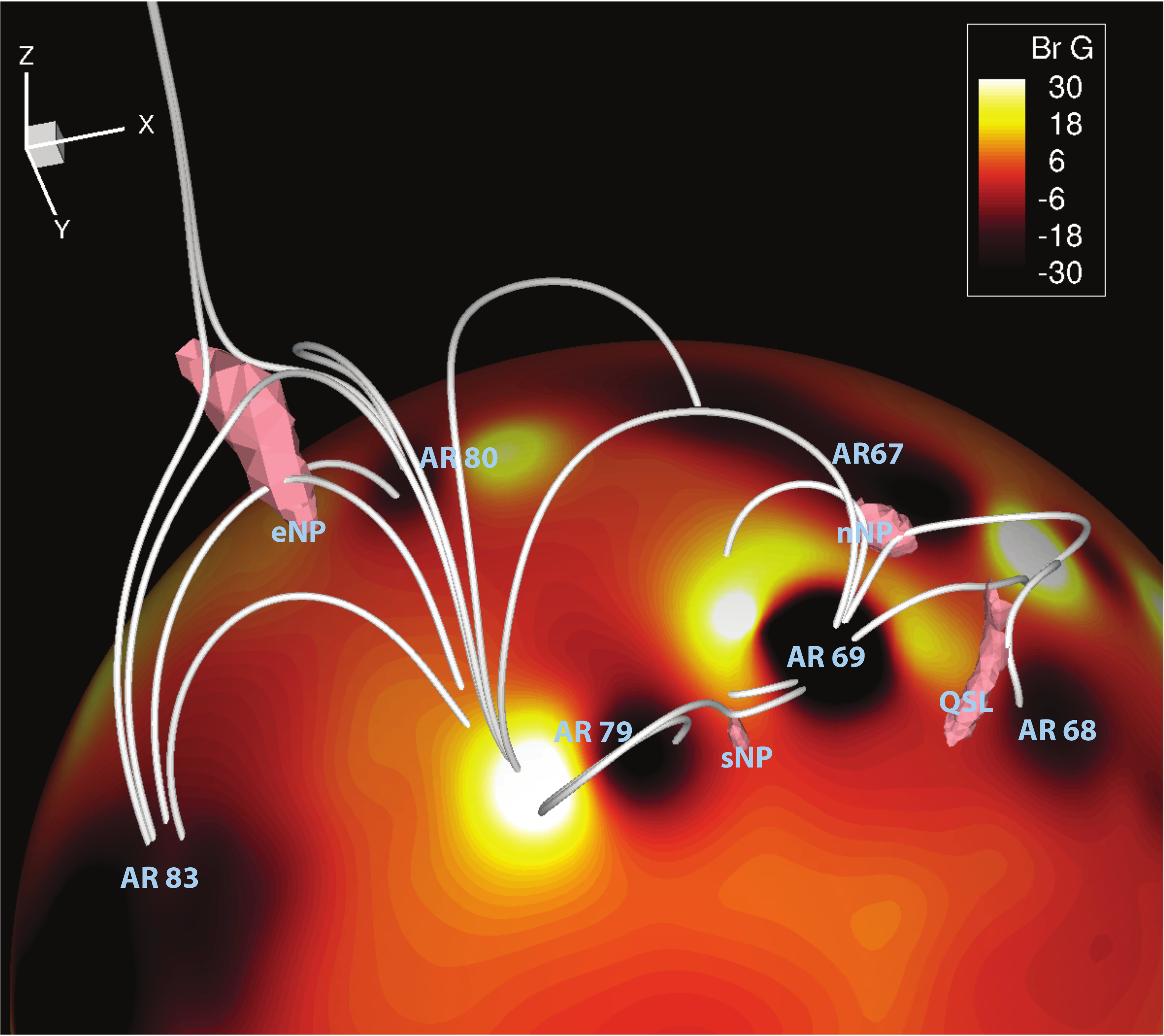}} 
{\includegraphics*[width=7.1cm]{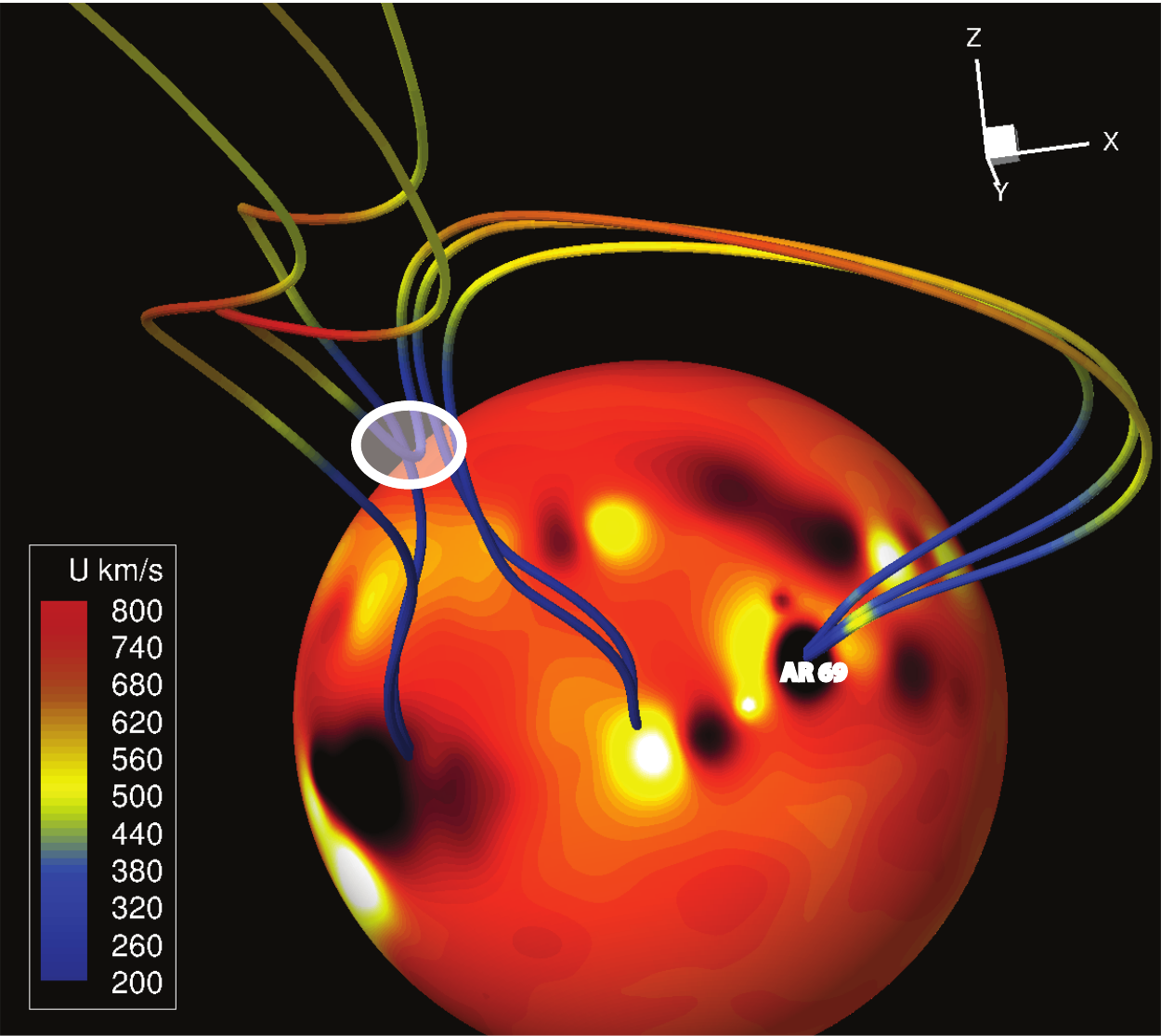}}
\end{center}
\end{minipage}\hfill
\caption{{\it Top}: Magnetic topology of the solar corona in the vicinity of AR 10069 on August 22-24, 2002. {\it Bottom}: Snapshot of the eruption at the end of the shearing phase (t = 1 hour) showing magnetic field lines color-coded with the radial velocity. The reconnection between the erupting flux from the dipole and the open magnetic field line happens at eNP and is highlighted with the white circle.} 
\end{figure}

Our simulation setup is similar to that discussed in more details in \citet{Roussev:2007}, \citet{Jacobs:2009} and \citet{Lugaz:2009a}.
In our numerical model, the steady-state solar corona and solar
wind are constructed following the methodology of \citet{Roussev:2003a}. The initial
condition for the coronal magnetic field is calculated by means of potential
field extrapolation with boundary conditions for
the radial magnetic field at the Sun provided by full-disk SoHO/MDI
observations. The plasma parameters are prescribed
in an ad-hoc manner, through a variable polytropic index, in order to mimic the
physical properties of streamers and coronal holes once a (non-potential) steady state is reached. 

To the initial magnetic field constrained by MDI data, we superimpose newly emerged magnetic flux
given by the dipolar magnetic field of two point charges, whose magnitude is chosen so that the peak value of the radial magnetic field 
at the solar surface is about 47~Gauss. To initiate the eruption, we use a similar strategy as described in \citet{Roussev:2007}.
To summarize, once the steady-state is reached at $t=0$, the two subsurface magnetic charges are
moved apart quasi-steadily up to $t=t_S=60$~min with a speed which is ramped up in $t_S/3$ to
V$_q$ = 33~km~s$^{-1}$; the charge motion is stopped at $t=t_S$. The two charges are moved approximatively parallel to the polarity inversion line. 
In order to provide sufficient buildup of magnetic energy, accompanying horizontal boundary motions are imposed as well.

\section{Results}
\subsection{Loss of Equilibrium}
One of the main results of the work in \citet{Roussev:2007} was to recognize the importance of the pre-existing magnetic topology in the initiation of the eruption. This is first and foremost 
because reconnection at the pre-existing null points (NP) and quasi-separatrix layers (QSLs) enables the sheared and energized magnetic flux of the dipole to erupt. 
There are four main pre-existing topological features important to understand these two eruptions (see top panel of Figure~1): a northern NP (nNP) between ARs 10067 and 10069 and a QSL
between ARs 10066, 10068 and 10069 for the August 24, 2002 eruption and two NPs for the August 22 eruption a southern one (sNP) between ARs 10069 and 10079 and an eastern one (eNP) between ARs 10079 and 10083.

For the August 24 eruption, more details can be found in \citet{Lugaz:2009a}. As a summary, the eruption is triggered when expanding field lines reconnect with the overlaying field through the QSL. As some of these field lines expand further, they reconnect though the nNP and some of them open up. The main motion of the erupting flux is radially above the initial position of the QSL but reconnection through the nNP enables
the expansion of the CME towards solar east, i.e. toward the Earth's direction. 

For the August 22 eruption, the EUV images show that the flare and CME occurred around the eastern side of AR 10069. Consequently, the two magnetic charges are put on the eastern side of AR 10069, approximatively midway between nNP and sNP. As the charges are moved, the initial reconnection, triggering the eruption, occurs at sNP and the eruption moves radially above this NP. Therefore, the initial direction of the August 22 eruption is to the south-east of the direction of the August 24 eruption, as also appears on the coronagraphic images. As the erupting field lines expand, they start to reconnect with open field lines at the eNP, resulting in magnetic connection between the source region of the eruption and the Earth (see bottom panel of Figure~1).

\subsection{Magnetic Connection to Earth}

At the start of the August 22 eruption, Earth was magnetically connected to the Sun through the eNP. The structure of this NP is such that the field lines in a cone 10$^\circ$ around Earth have footprints spanning from N20W50 to S15E10 on August 22. This explains why this S07W62 eruptions was very well connected to Earth and could result in a large SEP event. The flare site was only 20$^\circ$ away from open field lines connected to the vicinity of Earth. During the second half of the shearing phase, the erupting flux reconnects with some of these open magnetic field lines through the eNP (see bottom panel of Figure~1). This creates an opportunity for flare energetic particles to be further accelerated through the shock wave, creating a mix of flare- and shock-related particles (or more precisely of shocked superthermals and shocked flare particles).

\begin{figure}[t]
\begin{minipage}[]{1.0\linewidth}
\begin{center}
{\includegraphics*[width=7.1cm]{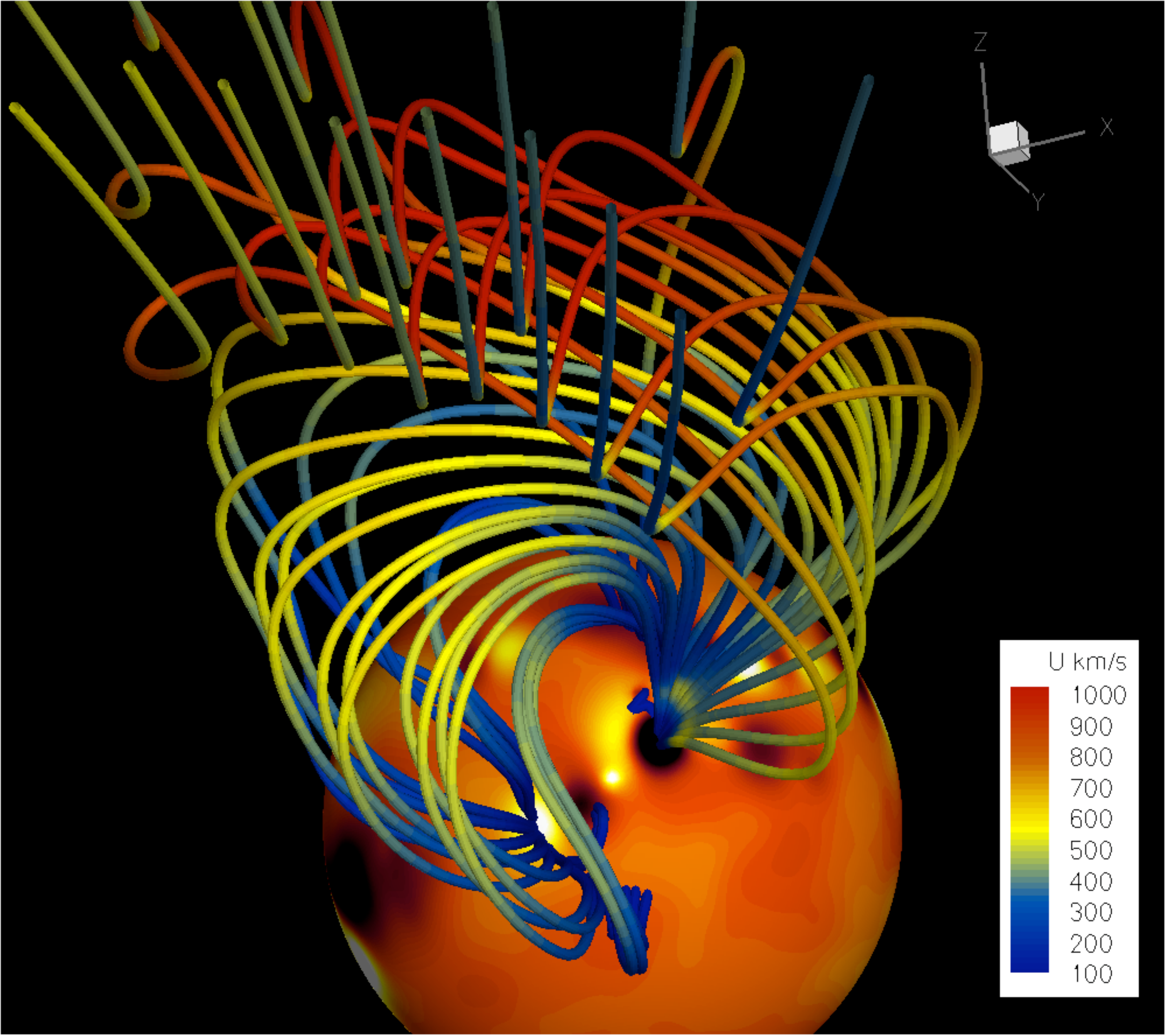}} 
{\includegraphics*[width=7.1cm]{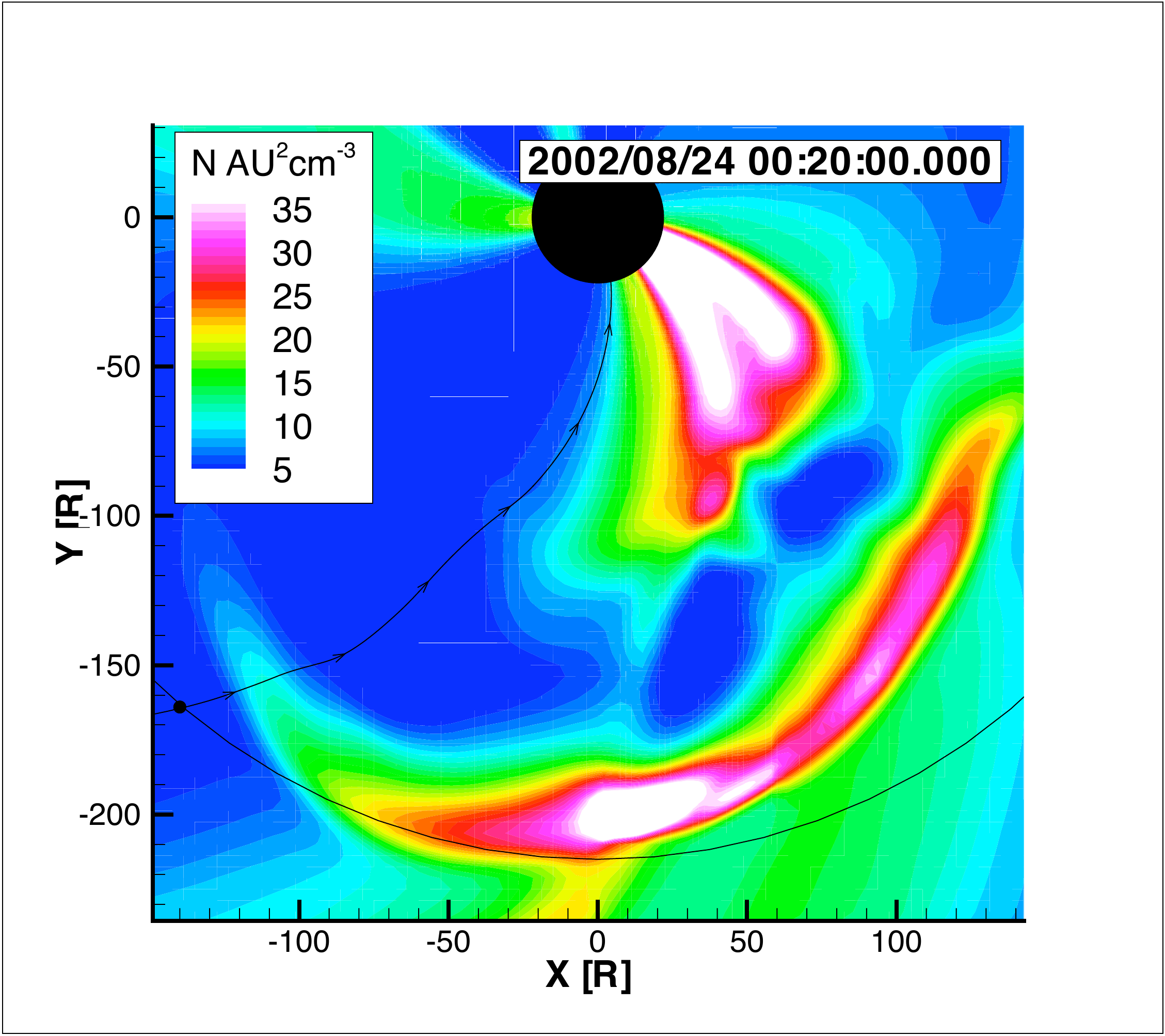}}
\end{center}
\end{minipage}\hfill
\caption{{\it Top}: Erupting magnetic field lines at the end of the shearing phase (t = 1 hour) color-coded with velocity. Note the absence of twist in the magnetic field. {\it Bottom:} Aug. 22 CME at the start time of the August 24 CME. The view is in the ecliptic plane, the large black disk represents the Sun, and the small one near (-140,-170) represents the Earth. The black circle has a radius of 1 AU. The contours show the density scaled by 1/r$^2$.}
\end{figure}

When we consider the magnetic connectivity between the source region of the ejection and the Earth at the start of the second eruption, we cannot ignore the presence of the previous ejection, which is at a distance of almost 0.95~AU (see bottom panel of Figure~2). Field lines connecting Earth and the Sun are still passing through the eNP, but they are modified due to the presence of a pressure disturbance associated with the previous eruption. We find that the length of the field line passing through Earth is significantly shorter when taking into account the August 22 CME compared to when it is calculated from the steady-sate solution (1.06 AU compared to 1.38 AU). This is because the field line goes through faster wind associated with the CME without being ``hindered'' by the closed field lines of the ejecta itself. This shorter length should be taken into account when deriving the release time of the particle as well as the height at which the shock wave associated with the August 24 eruption has formed. It should be noted that our steady-state solar wind is too fast compared to observations at 1 AU, and the precise values given here are therefore not reliable. 

\subsection{Heliospheric Evolution}

A shock wave was detected at Earth by the ACE spacecraft about 58~hours after the August 24 eruption. No shock wave was detected during the period August 19-25. One of the goals of our investigation is to determine whether it is possible that the August 22 eruption (W62) does not result in a shock wave detectable at Earth, whereas the August 24 eruption (W81) is associated with a shock detectable at Earth. As shown in \citet{Lugaz:2009a}, in our simulation, the shock wave associated with the August 24 eruption does not extend azimuthally all the way to Earth, even though it has an angular span of about 120$^\circ$. We also want to investigate whether or not the August 22 CME could be the cause of this shock wave. If it is not, the presence of a disturbed solar wind ahead of the August 24 CME, may explain at the same time the 58~hours transit time and the strong deceleration of the shock driven by the August 24 CME. As measured at 1~AU, the downstream region of the shock wave had a speed of about 450 km~s$^{-1}$ with a jump in speed of $\sim$ 50~km~s$^{-1}$ \citep[]{Tylka:2006}. Looking at the August 22 CME and its associated shock wave in our simulation (bottom panel of Figure~2), we find the opposite problem: the shock is detectable at 1~AU, 50 hours after the ejection, which shows that our CME model has too much azimuthal expansion. We should note that the solar wind speed is about 150 km~s$^{-1}$ too large before the arrival of the shock wave. Although this might partially explain the early arrival time, it is not enough to double it to match the observations.  It is also worth mentioning that the Empirical Shock Arrival (ESA) model \citep[]{Gopalswamy:2002, Xue:2005} for a CME whose initial speed is 1000 km~s$^{-1}$ (resp. 1900 km~s$^{-1}$) gives a transit  about 52 hours (resp. 24 hours). One of the few ways to associate the shock wave observed at Earth with one of these two CMEs, is to explain it as the flank of the shock wave driven by the August 24 CME, if its initial speed is 900~km~s$^{-1}$. This would imply a very wide and very curved shock front with a nose initially traveling at 2000~km~s$^{-1}$ and the flanks 90$^\circ$ away at 900~km~s$^{-1}$. In general, our simulations fail totally in explaining the arrival of a shock wave at Earth on April 26 associated with one of these two eruptions. The August 24 CME is not deflected enough and the August 22 CME cannot decelerate enough to arrive at Earth in over 4 days. Therefore, we look back at the CME list on August 22-24 to find a possible candidate. We believe the disk-center CME/flare (S4E22) on August 23 at 0548~UT is a good candidate for the driver of the shock observed on August 26. It appears to be associated with a partial halo whose speed is about 700-1000~km~s$^{-1}$. A 700~km~s$^{-1}$ CME is expected (ESA) to drive a shock which would arrive at Earth in about 74 hours, in good agreement with what is observed. 

\section{Discussions and Conclusion}
We have performed a Sun-to-Earth simulation of the August 22, 2002 CME event with a realistic CME initiation mechanism \citep[]{Roussev:2007}, following a similar simulation of the faster and more studied August 24, 2002 CME \citep[]{Lugaz:2009a}. We have found that, although the August 22 CME was not detected at Earth, it is likely that the presence of its associated shock wave and ejecta in the heliosphere significantly modified the length of the field lines connected to Earth at the start of the August 24 CME. This, in turn, must be taken into account when deriving the SEP release time and the shock formation height. We have also investigated the origin of the shock wave observed in situ at 1 AU on August 26 near 14UT. Our simulations are not able to reproduce the correct transit time and hit/miss from these two eruptions: the August 24, 2002 ejection is not wide enough to hit Earth in our simulation and the August 22, 2002 CME is predicted to hit Earth but 1.5 days earlier than what was detected. It is possible that complex interaction in the heliosphere between these two eruptions and other eruptions in the period from August 22-24, 2002 modified the transit time of one of the shocks to explain the observations at Earth \citep[as described in][]{Dasso:2009}. A disk-centered eruption on August 23 may play an important role in the formation of the detected shock. 

This study has shown the advantages and the limitations of modeling to study complex events such as fast, wide CMEs from the limb. The simulation can give us an estimate of the length of the field line connected to the Earth which can be, for example, compared to the ones computed from anisotropic measurements \citep[]{ERNE:2005} and must be used to derive SEP release time and the height at which the CME-driven shock wave was formed. Here, we have found that the length of the field line decreases by about 20$\%$. On the other hand, the model is not able to help to explain the arrival of a shock wave 2.5 days after the August 24 CME at a location about 80$^\circ$ east of the flare location. Although such Sun-Earth events originating from the western limb are rare, it will be important to better understand them in the future in order to have more reliable space weather prediction.


\begin{theacknowledgments}
  The research for this manuscript was supported by NSF grants ATM0639335 
  and ATM0819653 and NASA grants NNX07AC13G and NNX08AQ16G. 
  The SWMF was developed by the CSEM with funding support from NASA, NSF and DoD.
\end{theacknowledgments}



\bibliographystyle{aipprocl} 

\end{document}